 \documentclass[final,1p,times]{elsarticle}

 \usepackage{graphics}

\usepackage{amsmath}
\usepackage{amssymb}
\usepackage{subfigure}
\usepackage{float}

\usepackage{color}

\usepackage{rotating}
\usepackage{multirow}
\usepackage{comment}

\journal{J. Comput. Phys.}
\graphicspath{{./pics/}}
\newcommand \be{\begin{equation}}
\newcommand \ee{\end{equation}}
\newcommand \ba{\begin{eqnarray}}
\newcommand \ea{\end{eqnarray}}
\def\nn{\nonumber}

\begin{document}

\begin{frontmatter}



\title{
Spatially hybrid computations for streamer discharges with generic features of pulled fronts:
I. Planar fronts}


\cortext[cor1]{Corresponding author, Phone: 0031-20-5924208, Fax: 0031-20-5924200}

\author[CWI]{Chao Li\corref{cor1}}
\ead{Li@cwi.nl}
\author[CWI,TUE]{Ute Ebert}
\ead{Ebert@cwi.nl}
\author[CWI,RUN]{Willem Hundsdorfer}
\ead{Willem@cwi.nl}

\address[CWI]{Centre for Mathematics and Informatics (CWI), P.O.~Box~94079, 1090~GB Amsterdam, The
Netherlands}
\address[TUE]{Department of Applied Physics, Eindhoven University of Technology, P.O.~Box~513,
5600~MB, Eindhoven, The Netherlands}
\address[RUN]{Department of Science, Radboud University Nijmegen, Heyendaalseweg 135, 6525 AJ Nijmegen, The Netherlands}

\begin{abstract}

Streamers are the first stage of sparks and lightning; they grow due to a strongly enhanced electric field at their tips; this field is created by a thin curved space charge layer. These multiple scales are already challenging when the electrons are approximated by densities. However, electron density fluctuations in the leading edge of the front and non-thermal stretched tails of the electron energy distribution (as a cause of X-ray emissions) require a particle model to follow the electron motion. As super-particle methods create wrong statistics and numerical artifacts, modeling the individual electron dynamics in streamers is limited to early stages where the total electron number still is limited.

The method of choice is a hybrid computation in space where individual electrons are followed in the region of high electric field and low density while the bulk of the electrons is approximated by densities (or fluids). We here develop the hybrid coupling for planar fronts. First, to obtain a consistent flux at the interface between particle and fluid model in the hybrid computation, the widely used classical fluid model is replaced by an extended fluid model. 
Then the coupling algorithm and the numerical implementation of the spatially hybrid model are presented in detail, in particular, the position of the model interface and the construction of the buffer region. The method carries generic features of pulled fronts that can be applied to similar problems like large deviations in the leading edge of population fronts etc. 
\\


\end{abstract}


\end{frontmatter}

\newpage

\section{Introduction}\label{sec:intro}

\subsection{X-ray bursts and terrestrial gamma-ray flashes}

Terrestrial Gamma-Ray Flashes were first observed accidentally in 1994~\cite{Fis1994}; later they were found to be correlated with thunderstorm activity. In 2001, energetic radiation was also observed from normal cloud-to-ground lightning~\cite{Moo2001} and later from rocket triggered lightning~\cite{Dwy2003:2}. Meanwhile X-ray bursts were also found in the laboratory near long sparks~\cite{Dwy2005:2,Rahman2008,Ngu2008,Dwy2008}. 
 Such flashes and bursts are likely to be produced by Bremsstrahlung of very energetic electrons (so called run-away electrons).

One possible source of highly energetic electrons are the streamer discharges that pave the way of sparks and lightning. Streamers are ionized plasma channels that grow into a non-ionized medium due to the self-enhancement of the electric field at their tips. In this high field region, the electron energy distribution is very non-thermal and can have a long tail at high energies; it could act so much as a self-organized local electron accelerator that it is a possible source of X-rays~\cite{Mos2006,Li2007,Cha2008}.

The single electron dynamics in streamers can be studied by a particle model, which models the streamer dynamics at the lowest molecular level. It follows the free flight of single electrons in the local field between abundant neutral molecules and includes the elastic, exciting and ionizing collisions of electrons with the molecules in a stochastic Monte Carlo procedure. However, the increasing number of electrons in a growing streamer channel leads to an increasing demand of computational power and storage and excludes the simulation of all electrons in a full long streamer on present day's workstations. 

The generation of highly energetic run-away electrons from streamers therefore has only been modeled in particle models with simplifying assumptions or tricks. In~\cite{Mos2006} a Monte Carlo simulation treated planar fronts, and 3D aspects were modeled by a simplified electric field profile in the forward direction; in~\cite{Cha2008} a 3D axis-symmetrical streamer was simulated with super-particles of high weight where many real electrons are replaced by one super-particle. Both simulations indicate that electrons can be accelerated to very high energies in the high field zone at the tip of streamers. However, the 1D simulation~\cite{Mos2006} does not include the intricate 3D multiscale spatial structure of the streamer head properly~\cite{Dha1987,Vit1994,Ebe2006}, while the super-particle approach~\cite{Cha2008} has been shown~\cite{Li2008:2} to lead rapidly to numerical artifacts already shortly after a streamer emerges from an ionization avalanche; it will not represent the physically important tail of the electron energy distribution correctly.

\subsection{Need and concept of spatially hybrid computations for streamers}

Realizing that the number of electrons in the high field zone at the streamer tip is limited, and that only these electrons have a chance to gain high energies, a natural idea is to include only those electrons into a particle model while all others are treated in density or fluid approximation, as the fluid approximation is computationally much more efficient and has been widely used in streamer modeling, see e.g.,~\cite{Dha1987,Vit1994,Kul1994,Kul1995,Bab1996,Ebe1997,Arr2002,Roc2002,Mon2006:3,Seg2006,Luque2007,Luque2008:2,Luque2008:3,Nud2008,Eic2008}. Such a spatially hybrid approach should allow us to follow single electrons in the relevant region while avoiding the introduction of super-particles. We here present essential steps of its implementation.

\begin{figure}
\begin{center}
\includegraphics[width=0.60\textwidth]{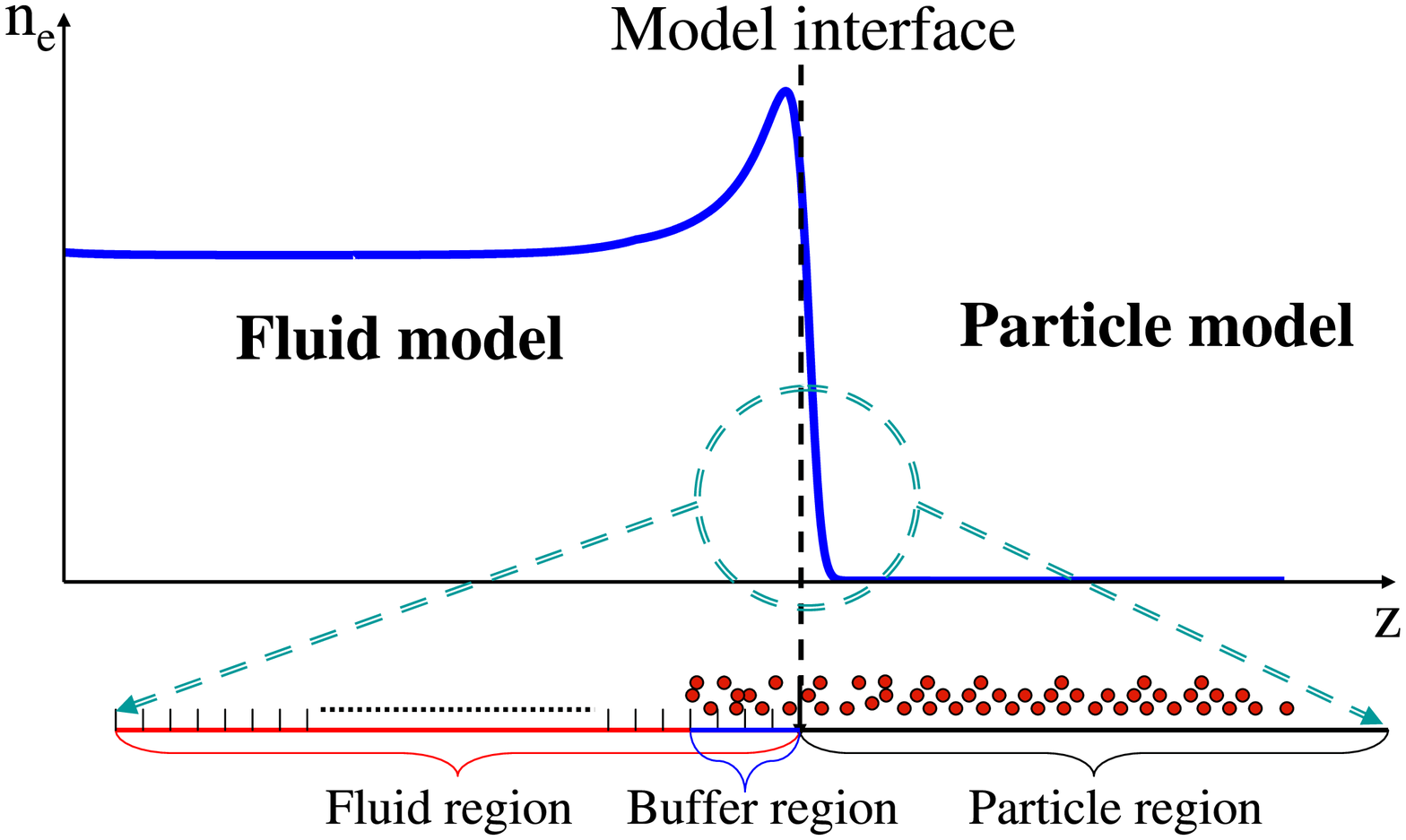}
\end{center}
\caption{\label{fig:hybrid_concept}(Color online)
 The spatially hybrid scheme is shown in a one dimensional section of the streamer. The solid line shows the electron density $n_e$ as a function of the spatial coordinate $z$. The particle model follows the leading part of the ionization front and the fluid model is applied to the rest. The questions treated in the present paper are: Which fluid model approximates the particle model in the best manner? Where should the model interface be placed and how should it be structured? The figure is reproduced from~\cite{Li2008:1}. }
\end{figure}

The concept of a spatial coupling of density and fluid model is shown in Fig.~\ref{fig:hybrid_concept}. The natural structure of the streamer consists of an ionized channel and an ionization front. In the channel, the electrons are dense enough to be approximated as continuous densities, and the field is too low to accelerate them to high energy. In the ionization front the electron density decreases rapidly while the field increases, and the electrons gain high energies far from equilibrium.
The hybrid model should apply the particle model in the most dynamic and exotic region with relatively few electrons and the fluid model in the remaining region that contains the vast majority of the electrons. The hybrid model is not only suitable for studying possible run-away electrons from streamer heads, but also for studying the role of electron density fluctuations on streamer branching, and for studying the gas chemistry inside the streamer and the electron energy distribution at the streamer tip in general.

To implement such a hybrid computation, we have set first steps in~\cite{Li2007,Li2008:1}. In~\cite{Li2007} we studied the Monte Carlo discharge model in detail and compared it with its fluid approximation; we did that for electron avalanches in a constant electric field and for planar ionization fronts; the study was for negative streamers in nitrogen. We derived reaction and transport coefficients for the fluid model from Monte Carlo electron avalanches, and we found that the mean electron energy increases at the front of an electron swarm, even if the electric field is constant. As the local field approximation cannot break down in a constant field, we attributed this effect to a break-down of the local density approximation of the fluid model in the steep density gradients of the advancing front. In the fast track comunication~\cite{Li2008:1}, we presented a first realization of a spatial coupling of particle and fluid model for planar ionization fronts. However, a closer investigation of these results indicates that further improvements are desirable, and the paper did not include details of our work. Therefore we here present details and further investigations on how to couple particle and fluid model in a setting of planar fronts. This lays the basis for full three dimensional streamer computations in the near future where the dynamics of single electrons in the relevant region can then be fully followed.

\subsection{Particle fluctuations and large deviations in pulled fronts} \label{sec13}

The hybrid coupling in space that we present here, is designed for the simulation of streamer discharges. However, the method carries generic features of so-called ``pulled fronts''. The pulling terminology was probably first used in 1976 in the biomathematical literature~\cite{Sto1976} for a phenomenon found in 1937 independently by Kolmogoroff, Petrovsky and Picounoff and by R.A. Fisher, examples include chemical~\cite{Dav1998}, bacterial~\cite{Gol1998}, and virus invasion fronts~\cite{For2002}. ``Pulling'' means that the penetration of one phase or species into another is dominated by the instability of the penetrated state; front shape and velocity are completely determined by the linear instability of the penetrated state, and not by a balance of forces of two competing states (for reviews see~\cite{Ebe2000:2,Saa2003}). Therefore, if the penetrating species is organized in particles (be it plants, bacteria, molecules or electrons), the penetration of one particle into the unstable state can create a local avalanche, and it can largely influence the dynamics. Large deviations from the mean in the leading edge of a front can therefore grow out into macroscopic structures. Examples are the first electron that reaches a completely non-ionized region with a local field above the ionization threshold, or the first species that reaches a new habitat~\cite{Edm2004,Klo2006,Hal2008}. In that sense, it is surprising how well fluid approximations work for many pulled fronts.

Spatially hybrid models therefore are of general interest and a viable method for pulled front problems as they treat the dynamically relevant leading edge with low particle densities in a particle model and the bulk of particles behind the front in density approximation. 
In particular, for off-lattice models it poses a challenge. Questions include: Which fluid model approximates the particle model in the best manner? Where should the model interface be placed and how should it be structured?

Adapting the buffer zone method~\cite{Gar1999,Ale2002,Ale2005,Del2003,Akt2002} to connect the particle with the fluid zone, we encounter two more features that are generic for pulled fronts. 1. When the model interface moves along with the velocity of the front, the particles in this frame of reference move backward. This is because the front is pulled by motion and multiplication of particles in the leading edge where the electric field and the electron drift velocity are the highest, therefore throughout the front, individual particles on average move backward, even in the leading edge. As we will show in Section~\ref{sec:buffer_region}, this leads to a large technical advantage for the construction of the buffer region, as particles do not need to be created out of the fluid region, but they only have to be absorbed there. 2. Avalanches in the unstable region have essentially the same front shape and velocity as full fronts~\cite{Ebe2000:2,Li2007}. When particle avalanches are used to derive transport and reaction functions for the fluid approximation, a pulled front is therefore well matched by this model.

\subsection{Content of the paper}

Section~\ref{sec:p_f_models} discusses why our first hybrid coupling approach~\cite{Li2008:1} still showed discontinuities at the model interface. The answer is found in inconsistent fluxes, and the problem is resolved by improving the fluid model through a gradient expansion in the density. Then transport and reaction coefficients are derived for this extended fluid model, and the extended fluid model is compared to the full particle model, both for avalanches and for planar fronts. The extended fluid model now describes the dynamics well, except for particle fluctuations in the leading edge of the front.
The coupling of the extended fluid model with the particle model is described in Section~\ref{sec:h_model}, where the coupling concept and implementation are given in detail; discussed are the hybrid algorithm, the numerical implementation, the position of the model interface and the structure of the buffer region at the interface. While this whole study is done in a fixed background field for negative streamer fronts in nitrogen, Section~\ref{sec:Ch6_other_field} investigates how these results are generalized to other fields. Section~\ref{sec:ch6_conclusion} contains the conclusions. 
Appendix~\ref{sec:Energy_equation} discusses an alternative to the extension of the fluid model by a gradient expansion, namely the extension with an electron energy equation.

\section{Extending the fluid approximation}\label{sec:p_f_models}

The particle model follows the individual electrons on their deterministic flight in the local field until the next collision with a neutral particle. Time and kind of collision and energy and momentum of the electron after the collision are calculated in a Monte Carlo procedure, in accordance with the cross-section data base of~\cite{Siglo}. Details of the procedure are explained in ~\cite{Li2007}.

The classical fluid model as used by many authors for negative streamers~\cite{Dha1987,Vit1994,Kul1994,Kul1995,Bab1996,Ebe1997,Arr2002,Roc2002,Mon2006:3,Luque2008:3} can be derived from the particle dynamics through the Boltzmann equation assuming the local field and the local density approximation, where the electron mean energies, transport coefficients and reaction rates are functions of the local reduced electric field and the local electron density. (We remark that the local field approximation is typically emphasized in this derivation, while the local density approximation is not.)
However, we have shown that in electron avalanches, the local mean energies of the electrons vary from the tail to the front of the swarm~\cite{Li2007}, even though the electric field is constant. It is impossible to improve the fluid model to the same level of description as the particle model, and we do not intend to do that. But the main features that create inconsistencies and that influence the coupling of the two models, have to be investigated and understood.

Here we investigate why the electron fluxes are inconsistent when coupling the classical fluid model with the particle model, and we extend the fluid model with a gradient expansion in the electron density to incorporate nonlocal effects. The two fluid models are compared with each other for both the electron avalanche and the planar front.

\subsection{ A short recollection of the classical fluid model}

To elucidate the essential structure of the problem, we here analyze discharges in pure nitrogen.
The classical fluid model then contains two continuity equations for the densities $n_e$, $n_p$ of electrons and positive ions
\ba \frac{\partial n_e}{\partial t} + {\bf \nabla}\cdot {\bf j}_e &=& {\cal S}_l, \label{ch6_eq:fluid1}\\
\frac{\partial n_p}{\partial t} \hspace{1.0cm} &=& {\cal S}_l, \label{ch6_eq:fluid2} \\
{\bf j}_e &=& - \mu_l(E) {\bf E} n_e - {\bf D}_l({\bf E}) \cdot{\bf \nabla} n_e, \label{ch6_eq:flux} \ea
where ${\bf j}_e$ is the electron flux and ${\cal S}$ is the source of electrons and ions, $\mu_l$ represents the mobility and ${\bf D}_l$ is the diffusion matrix, $\bf E$ and $E$ are the electric field and the electric field strength; the subscript ``$l$'' stands for ``local'' and denotes the classical fluid model in contrast to the extended one to be introduced below. The densities of charged particles together with the boundary conditions determine the electric field through the Poisson equation
\begin{equation}
\nabla \cdot{\bf E}= \frac{{\rm e}\;(n_p-n_e)}{\epsilon_0} \label{ch6_eq:Poisson}.
\end{equation}
The source term accounts for the impact ionization in local field and local density approximation
\be
\label{ch6_eq:Sl} {\cal S}_l=|n_e\;\mu_l(E)\;{\bf E}|\;\alpha_l(E) . \ee

\subsection{Why the fluid model needs an extension}\label{sec:why_ex}

\subsubsection{Discontinuities in the hybrid model for planar fronts}

\begin{figure}
\begin{center}
\includegraphics[width=0.6\textwidth]{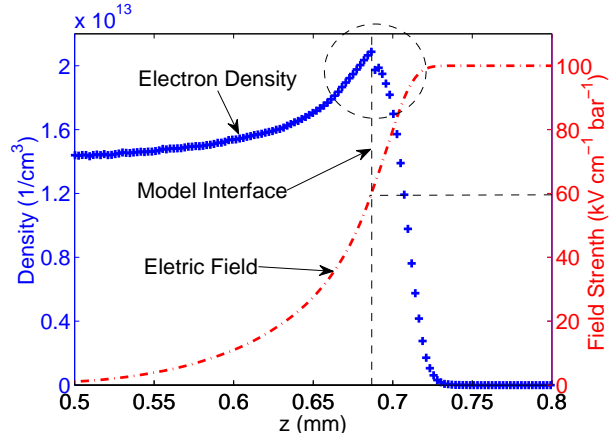}
\end{center}
\caption{\label{fig:discontinuity} Hybrid simulation of a planar front propagating in $E^+=-100$ {\rm kV/cm} with the model interface at -60 {\rm kV/cm}. The hybrid model introduced in~\cite{Li2008:2} is used. Electron density (``+'') and electric field (dot-dashed line) are plotted, and a discontinuity in the electron density is marked with a dashed circle.}
\end{figure}

We have derived transport and reaction coefficients for the classical fluid model from the particle model in~\cite{Li2007}, and we have shown first results of a hybrid computation which couples the classical fluid model with the particle model in~\cite{Li2008:1}. 

However, to ensure a stable and correct interaction between the two models, we later investigated the electron flux densities on the model interface, and the mean electron energies and velocities around the model interface, and we found a disagreement of the local electron flux between the two models. When the particle and the classical fluid model are applied in the same region of the ionization front, the mean electron flux density is lower in the particle simulation than in the classical fluid simulation as we will demonstrate below. If one places the model interface near the maximum of the electron density, a density jump can appear near the model interface. This is shown in Fig.~\ref{fig:discontinuity} where the model interface is placed at the electric field $E = -60$ kV/cm with the field ahead of the front being $E^+= -100$ kV/cm; here the local fluxes are discontinuous across the model interface, and this results in the visible discontinuity of the density. The numerical discretization and other details of this experiment can be found in~\cite{Li2008:1}.

The effect appears although the transport and reaction coefficients were generated in the particle swarm experiments and then implemented into the fluid model; therefore the two models should be consistent. To understand the underlying problem, we review the conceptual differences between the two models and compare them in a electron avalanche experiment.

\subsubsection{Electron avalanche simulations in particle and fluid model - a reinvestigation}\label{sec:Swarm_reinvestigation}

\begin{figure}
\begin{center}
\includegraphics[width=0.65\textwidth]{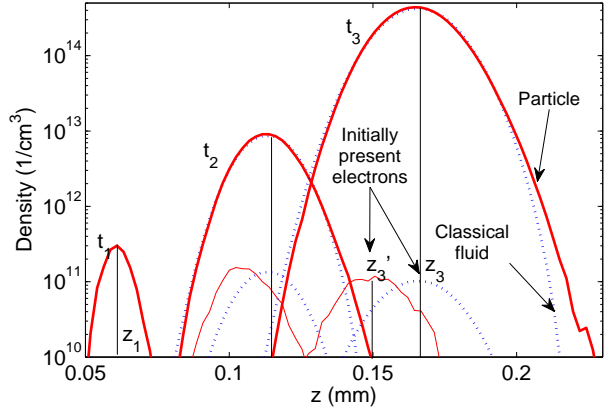}
\end{center}
\caption{\label{fig:swarm_com_1} Evolution of electron swarms at three consecutive time steps $t_1$, $t_2$, $t_3$ in a field of $-100$ {\rm kV/cm}, where $t_1$, $t_2$, and $t_3$ are $20$, $130$, $240$ {\rm ps} when the simulation starts from 100 pairs of electrons and ions. Shown are the spatial profiles of the total swarm in the 1D particle model (solid line) and in the 1D classical fluid model (dotted line). Also shown are the spatial profiles of those electrons that were already present at time $t_1$ (at the same three time steps). }
\end{figure}

Fig.~\ref{fig:swarm_com_1} shows simulated electron swarms propagating in a uniform constant field of
$-100$~kV/cm. The simulations have been carried out both with the particle model (solid lines) and with the
classical fluid model (dotted lines) in 1D starting from the same initial conditions. For the particle model, 1D means that the simulations have been carried out in a volume with a narrow transversal cross-section, with periodic boundary conditions at the lateral edges.

Here and in all later comparisons of particle or fluid results, the initial distribution of electrons and
ions is generated in the following manner. An electrically neutral group of 100 electrons and ions is
inserted at one point in space. Their temporal evolution is calculated with the particle model. After a
short time that depends on the electric field, e.g., after $t_1=20$ ps in a field of 100~kV/cm, a small swarm 
forms in which the electron density profile is well approximated by a Gaussian distribution. This
distribution of electrons and ions is used as an initial condition for all simulations, including particle, fluid and hybrid.

In the particle model, the standard Monte Carlo technique in a constant electric field is applied. 
For the fluid model, the equations of the particle densities are
discretized in space with a finite volume method. The particle densities are updated in time using a third
order upwind-biased advection scheme combined with a two-stage Runge-Kutta method. The same time step and
cell size were used in the particle and the classical fluid calculation, i.e., $\Delta t= 0.3$ ps and
$\Delta z= 2.3$ $\mu$m. More details of the numerical implementation of particle and fluid model can be
found in~\cite{Li2007}.

In the avalanche experiment, we let both the particle model and the classical fluid model follow the swarm from
$t_1=20$ ps to $t_3=240$ ps, and the electrons in the particle simulation are mapped to densities on the 
same grid as used in the fluid simulation. During the simulation, we follow the growth of the 
total electron avalanche, and we pay specially attention to the group of electrons that are 
initially (at time $t_1$) present as seed electrons.
Fig.~\ref{fig:swarm_com_1} shows the electron density profiles of the swarm at time $t_1$ and at the later
stages $t_2$ and $t_3$, and also the density distributions of only those electrons that already existed at time $t_1$, at all three time steps $t_1$, $t_2$ and $t_3$, neglecting all electrons generated later. 

Fig.~\ref{fig:swarm_com_1} shows that the profiles of the total swarm are nearly the same in the classical
fluid model and in the particle model. This was to be expected as we just derived the transport and reaction
coefficients $\mu_l$, ${\bf D}_l$, and $\alpha_l$ for the fluid model through this condition~\cite{Li2007}. 
However, following only the swarm 
of electrons that already existed at time $t_1$, the particle swarm stays behind the fluid swarm.
This observation has to be interpreted in the following manner.
In the fluid simulation, the electron swarm grows homogeneously, i.e.,
the center of mass of the initially present electrons moves with the same speed as the center of the mass of
total electron swarm. However, in the particle simulation the mean displacement of the whole swarm is larger
than that of the initially present particles. This effect can only be caused by a larger growth rate 
of the electron swarm at the tip than at the back of the swarm.
This shows that the classical fluid model is based on approximations that here become inaccurate.

The electron flux rates in particle or fluid model can be derived by averaging 
the local electron flux ${\bf j}_e$ over the simulation domain of volume $V$ and over a
short time interval $\tau$
\begin{equation}\label{EQ-flux}
\frac{\bar{\bf j}_e}{N_e}=  \int_t^{t+\tau} \frac{dt}\tau \int_V  dV\; \frac{{\bf j}_e}{ N_e},
\end{equation}
where $N_e$ is the total number of electrons. We find a higher mean flux density per electron
in the fluid than in the particle swarm simulation. The mean flux density per electron equals 
the mean electron drift $\mu {\bf E}$, as the diffusive flux vanishes if all electrons are inside 
the integration volume, see Eq.~(\ref{ch6_eq:flux}). The higher electron mobility in the fluid model is consistent with the larger displacement of the swarm of initially present electrons.

We conclude that the particle fluxes are inconsistent between particle model and classical fluid 
approximation both for planar fronts and for electron avalanches. For the avalanches, this has been 
shown both by interpretation of Fig.~\ref{fig:swarm_com_1} and by a calculation of fluxes through 
Eq.~(\ref{EQ-flux}).

\subsubsection{Discussion: validity of the local density approximation}

 That both the mean flux density (\ref{EQ-flux}) and the avalanche growth characteristics 
(Fig.~\ref{fig:swarm_com_1}) differ between particle model and classical fluid approximation, is 
actually due to the same reason: the local density approximation. As the electric field is constant 
in the swarm experiment, it cannot be the local field approximation.

In~\cite{Li2007}, it was already shown that the local electron energy distribution depends
on the local electron density gradients both in planar fronts and in electron avalanche. It indicates that
even in a uniform constant field, the ionization rates increase from the tail to the head of the electron
swarm. Therefore there are two contributions to the speed of the swarm: i) the electron drift in the local
field as a main contribution and ii) the unequal growth of the swarm due to the unequal electron energy
distribution.

The classical fluid model assumes that the local mean energies and local mean reaction rates are functions
of the reduced electric field, and it ignores the fact that within the same electric field, the ionization rate
can be different. At the front of the swarm, this leads to a lower reaction rate in the fluid simulation
than in the particle model. But by overestimating the ionization rate at the tail, the swarm in the
classical fluid model generates the same amount of electrons as in the particle swarm. And by overestimating the
mobility of electrons, the swarm nevertheless propagates with the same speed. Or in other words, the
classical fluid model approximates the swarm density profile rather well, but at the price of wrong local
fluxes and wrong local reaction rates. In simulations with non-uniform fields, such as streamers in which
the density profiles are determined only by the front while the tail doesn't play a role in the propagation, the classical fluid model pays the price by generating lower electron densities inside the streamer 
than the particle model~\cite{Li2007}.

Therefore the fluid model has to be extended to include the dependence of the ionization rate on
non-uniform electron density conditions. And by introducing a nonlocal term, we expect the fluid model to
supply the same electron flux as the particle model within the same field.

\subsection{Gradient expansion and extended fluid model}\label{sec:f_equations}

\subsubsection{The introduction of the extended fluid model}

The electron ionization rates for non-uniform fields and electron densities are discussed
in~\cite{Ale1996,Nai1997}. There perturbation theory is used to obtain the rate coefficients, and it is
shown that they can be represented by a gradient expansion about the local values
\begin{equation}
{\cal S}={\cal S}_l\;\Big(1+k_1 \hat{\bf e}\cdot\nabla \ln n_e +k_2 N/E \nabla \cdot ({\bf{E}}/N) + k_3 \hat{\bf e}\cdot\nabla \ln (E/N)\Big), \label{ch6_eq:Ale}
\end{equation}
where $n_e$ is the electron density, $\bf{E}$ and $E$ are the field and the field strength and $\hat{\bf e}={\bf E}/E$ is the unit vector in the direction of the electric field, $N$ is the
molecule density of the background gas, and $k_0$, $k_1$, $k_2$, $k_3$ are
parameters depending on $E/N$ that have to be determined.

In~\cite{Li2007}, we have shown that the local density approximation is
insufficient while the local field approximation is not problematic. Therefore, we neglect the field
gradient terms, focus on the density gradient in Eq.~(\ref{ch6_eq:Ale}), and approximate the source term in our nonlocal model for fixed density $N$ as 
\ba
\label{ch6_eq:S} {\cal S}_n&=&{\cal S}_l \;\Big(1+k_1(E) \;\hat{\bf e}\cdot\nabla \ln n_e \Big)
\nn\\
&=& \mu_n(E)\;\alpha_n(E)\;\Big( E\; n_e + k_1(E)\;{\bf E}\cdot\nabla n_e\Big), 
\ea
where the subscript $n$ denotes the nonlocal or extended fluid model. Otherwise the model is identical to the classical fluid model, except that all transport and reaction coefficients have to be determined a new, they are denoted as $\mu_n$, ${\bf D}_n$, and $\alpha_n$. All these functions as well as $k_1$ are now determined.

It has been known for long that the consistency of the fluid model with particle swarm results is important. Robson {\it et al.}~\cite{Rob2005} recently reviewed this issue and emphasized that the consistency requirement actually applies to transport coefficients and reaction rates. 
In~\cite{Li2007}, we therefore have determined the mobility $\mu_l({E})$, diffusion tensor ${\bf D}_l({E})$, and ionization rate $\alpha_l(E)$ for the classical fluid model from particle swarm experiments.

\subsubsection{Electron mobility $\mu_n$}

When the classical fluid model is extended with a density gradient term, the transport and ionization
coefficients may also change and should be re-defined from the swarm experiments. As discussed above, the mobilities can be determined either i) from the mean displacement of electron swarm in a uniform constant field \[\mu_l(E)E= \frac{z_3-z_1}{t_3-t_1},\] (as was done in the classical fluid model) or ii) from the mean displacement of the initially present particles
\[\mu_n(E)E= \frac{z_3'-z_1}{t_3-t_1},\]
where $z_1$ and $z_3$ are the centers of mass of the swarms at times $t_1$ and
$t_3$, and $z_3'$ is the center of mass of those electrons at time $t_3$ that were already present at time $t_1$, cf.~Fig.~\ref{fig:swarm_com_1}.
The second definition represents the real electron motion and is used in the extended fluid model; within numerical accuracy, this mobility coincides with the one derived from the average local flux rate (\ref{EQ-flux}). On the contrary, the first definition is actually the sum of the local electron fluxes under the influence of the electric field plus the nonlocal growth of the swarm.

\begin{figure}
\begin{center}
\includegraphics[width=0.6\textwidth]{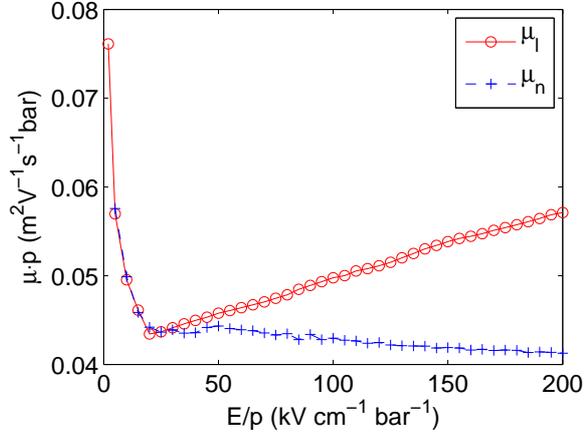}
\end{center}
\caption{\label{fig:mus} $\mu_l$ and $\mu_n$ as functions of the electric field strength $E$.}
\end{figure}

The mobilities
$\mu_l$ or $\mu_n$ are plotted in Fig.~\ref{fig:mus}. Here $\mu_l$ is the electron mobility in the
classical fluid model and $\mu_n$ is the one in the extended fluid model. They are almost the same for
fields $E$ below 30~kV/cm when impact ionization is small or even completely negligible, but the difference
increases as the field strength increases. 
It can be noted that the curves in Fig.~\ref{fig:mus} are not
completely smooth. This is due to stochastic fluctuations within the Monte Carlo simulations. However,
it has been checked that these small fluctuations have a negligible influence on the final results.

\subsubsection{Impact ionization rate $\alpha_n$}

The reaction rate $\alpha$ is derived from the growth rate of the total electron number $N_e(t)$ in a particle swarm simulation. Within a time $t$, the swarm propagates a distance $\mu(E)E t$, and the reaction rate is determined by \be \alpha(E) = \frac{\ln N_e(t_2)-\ln N_e(t_1)}{ \mu(E) E\;(t_2-t_1)}. \ee
When $\mu_l$ is replaced by $\mu_n$, the ionization rate changes from $\alpha_l$ to \be
\alpha_n(E) = \alpha_l(E)\;\frac{\mu_l(E)}{\mu_n(E)}\ee as the simulation fixes the product $\mu_l\alpha_l=\mu_n\alpha_n$.

\subsubsection{Diffusion tensor ${\bf D}_n$}

The diffusion rates can also be obtained according to two different definitions: the diffusion of the
electron swarms ${\bf D}_l$ or diffusion of the initially present electrons ${\bf D}_n$. The results turn out to be the same
\be \label{Eq:D}
{\bf D}_l({\bf E})={\bf D}_n({\bf E}),
\ee
as shown below.

\subsubsection{The parameter function $k_1$ of the gradient expansion}

The extended fluid model needs the function $k_1$ in Eq.~(\ref{ch6_eq:S}). The rates $k_1$, $k_2$, and $k_3$ in Eq.~(\ref{ch6_eq:Ale}) were calculated by Aleksandrov and Kochetov~\cite{Ale1996} by solving the Boltzmann equation in a two-term approximation. But to stay consistent with the approach above, all parameter functions in the fluid model are here derived from particle swarm simulations. Since we neglect the field gradient term in the source term, only $k_1$ needs to be calculated.

\begin{figure}
\begin{center}
\includegraphics[width=0.6\textwidth]{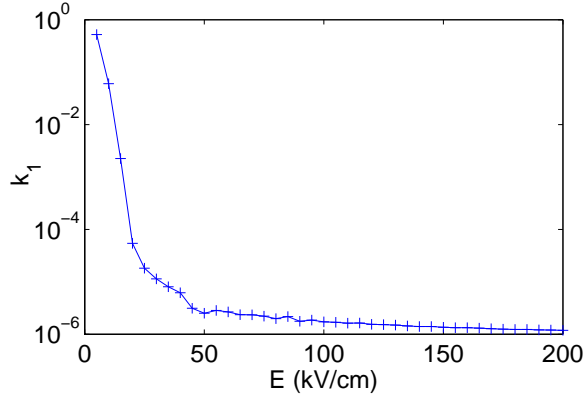}
\end{center}
\caption{\label{fig:k_1} The parameter function $k_1$ as a function of the electric field.}
\end{figure}

In fact, $k_1$ can be calculated by simply comparing the electron continuum equations in both the classical
and the extended fluid model. As both fluid models can properly describe the swarm when appropriate parameters are applied, we have
\ba \partial_tn_e&=&\nabla \cdot\Big(\mu_l(E) {\bf E} n_e\Big)   +  \nabla\cdot\Big( {\bf D}_l({\bf E}) \cdot{\bf \nabla} n_e\Big)  +   n_e\;\mu_l(E)\;E\;\alpha_l(E)  \\
&=& \nabla \cdot\Big(\mu_n(E) {\bf E} n_e\Big)  +  \nabla\cdot\Big( {\bf D}_n({\bf E})\cdot {\bf \nabla} n_e\Big)  + n_e\;\mu_n(E)\;E\;\alpha_n(E)
+k_1(E) \;\mu_n(E)\;\alpha_n(E)\;{\bf E}\cdot\nabla n_e.\nonumber
\ea
In a constant uniform electric field such as in the electron avalanche experiment, $\mu_l(E)$, $\mu_n(E)$ and $E$ are constant and $\nabla\cdot\Big( \mu_l({\bf E}) \;{\bf E}\; n_e\Big)= \mu_l({\bf E}) \;{\bf E}\cdot\nabla n_e$ etc. Removing identical terms on both sides of the equation, we retrieve ${\bf D}_l={\bf D}_n$ (\ref{Eq:D}) and we get $\mu_l(E)=\mu_n(E) ( 1 + k_1(E) \alpha_n(E) )$ or
\be   k_1(E) =\frac{\mu_l(E)-\mu_n(E)}{\alpha_n(E)\;\mu_n(E)} \label{eq:ch6_k1_cal}
\ee
where $\mu_l$, $\mu_n$ and $\alpha_n$ were derived above from the particle swarm experiments. $k_1$ as a function of the field is shown in Fig~\ref{fig:k_1}.

\subsection{Comparison of the extended fluid model with the particle model}\label{sec:com_ef_mc}

Now the stage is set to compare the extended fluid model with the particle model both in swarms and in planar ionization fronts. The extended fluid equations are discretized in the same manner as the classical fluid model, and the particle densities are updated with the same scheme as before. The same time step and cell size are used in the extended fluid calculation as in the particle and the classical fluid calculation (see Sect.~\ref{sec:why_ex}).

\subsubsection{Swarm simulations}

\begin{figure}
\begin{center}
\includegraphics[width=0.65\textwidth]{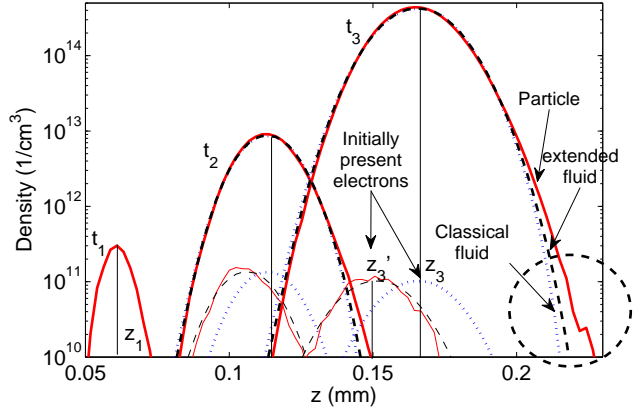}
\end{center}
\caption{\label{fig:swarm_com} The density profiles of electron swarms and the initial electrons in a field of $-100$~kV/cm are shown at times $t_1$, $t_2$ and $t_3$. The plot is the same as in Fig.~\ref{fig:swarm_com_1}, but now the extended fluid simulation is added as a dashed line.}
\end{figure}

In Fig~\ref{fig:swarm_com}, we show electron swarms at times $t_1$, $t_2$, and $t_3$ in a constant field of $-100$~kV/cm; the swarms were followed by particle simulation (solid line), classical fluid simulation (dotted line) and extended fluid simulation (dashed line). For the whole swarm, all three models give similar results, but the extended fluid model follows the evolution of the initially present electrons much better than the classical fluid model.

However, the figure also shows that in the leading edge of the swarm (marked with a large dashed circle) neither the extended fluid model nor the classical fluid model describe the electron density distribution of the particle model precisely. As described in~\cite{Li2007}, individual electrons with high energies largely deviate from the mean, and none of the two macroscopic fluid models can reproduce this microscopic behavior. The densities in this region are 3 to 5 orders lower than the maximal density.
But as streamer ionization fronts are so-called ``pulled'' fronts, cf.~section~\ref{sec13}, the behavior in the leading edge of the front actually determines the front velocity. From this point of view, the hybrid model that uses the particle model in the leading edge of the ionization front, is the method of choice.

\begin{figure}
\begin{center}
\includegraphics[width=0.6\textwidth]{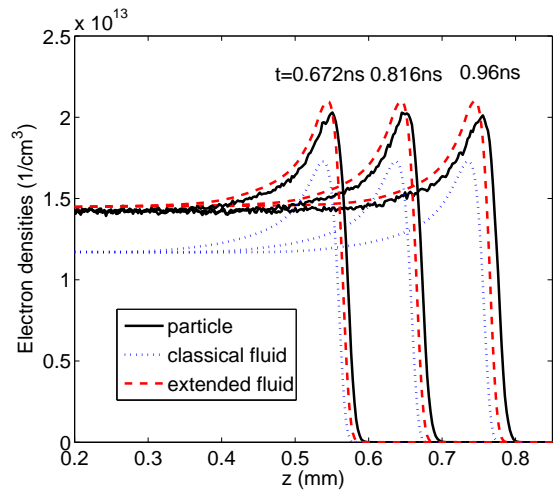}
\end{center}
\caption{\label{fig:den_com_100} Temporal evolution of the electron densities in a planar front in a field of $E^+=-100$ kV/cm. Shown are the spatial profiles of electron densities derived with the particle, classical fluid or the extended fluid model at time levels t=0.672 ns, 0.816 ns and 0.96 ns (particle simulation: solid line, fluid simulation: dotted line, and extended fluid simulation: dashed line).}
\end{figure}

\subsubsection{Front simulations}

Fig.~\ref{fig:den_com_100} shows the temporal evolution of the planar front in a field of  $E^+=-100$ kV/cm in the particle simulation (solid line), classical fluid simulation (dotted line), and extended fluid simulation (dashed line). Compared to the classical fluid model, where the maximal electron density in the front and the saturation level of the ionization behind the front are about $20\%$ lower than in the particle model, the extended fluid model approximates the particle model much better. The particle and fluid front move with approximately the same velocity, but the particle front moves slightly faster than the extended fluid front, and the extended fluid front moves slightly faster than the classical fluid front, in agreement with Fig.~\ref{fig:swarm_com}. (We recall from~\cite{Li2007} that the leading edge of a swarm and of a pulled front in the same field have the same spatial profile and create the same velocity.)

\begin{figure}
\begin{center}
\includegraphics[width=0.6\textwidth]{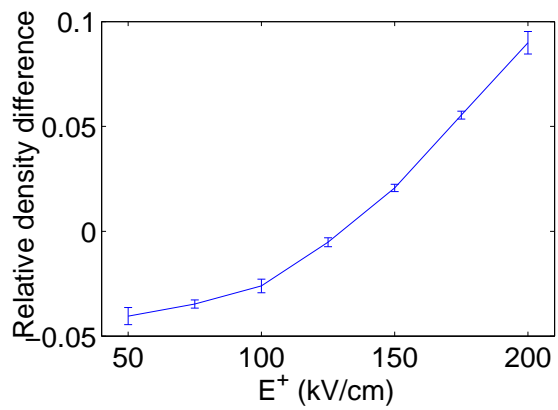}
\end{center}
\caption{\label{fig:nl_mc} The relative density difference (\ref{EQ-dens}) between extended fluid model and particle model.}
\end{figure}

Having analyzed planar fronts at $-100$~kV/cm, we now summarize the front results for fields ranging from $-50$ to $-200$~kV/cm in Fig.~\ref{fig:nl_mc}. The figure shows the relative difference
\be \frac{n^-_{e,part}-n^-_{e,e.fluid}}{n^-_{e,part}}. \label{EQ-dens}\ee
of the saturated electron density behind the front in the particle model ($n^-_{e,part}$) and the extended fluid model ($n^-_{e,e.fluid}$).
Compared to the relative difference between the particle model and the classical fluid model discussed in~\cite{Li2007}, which increases from $10\%$ at 50~kV/cm to $40\%$ at 200~kV/cm, we now find that the relative density difference in the extended fluid model never exceeds $10\%$ within this range of fields.

\subsection{The drawbacks of the extended fluid model}\label{sec:drawback}

By approximating the nonlocal ionization rate with a density gradient expansion, the extended fluid model reproduces the ionization level behind the front much better than the classical fluid model. However, Fig.~\ref{fig:nl_mc} shows that this ionization level in the extended fluid model exceeds the one in the particle model when the field $E$ is below 125~kV/cm. This will not harm its coupling with the particle model, but the reason for this unexpected behavior is briefly discussed here, because it could lead to further improvements of the fluid model.

Several possible reasons have been examined, for example, the quality of the parameters ($\mu$, {\bf D}, $\alpha$ and $k_1$) used in the extended fluid model, the discretization of the gradient expansion term, and the relative ionization rate in the very leading edge where the electron density simply vanishes in the particle model, but remains nonzero in the fluid model due to electron diffusion. The answer was found by comparing the ionization rates within the front between both models.

\begin{figure}
\begin{center}
\includegraphics[width=0.6\textwidth]{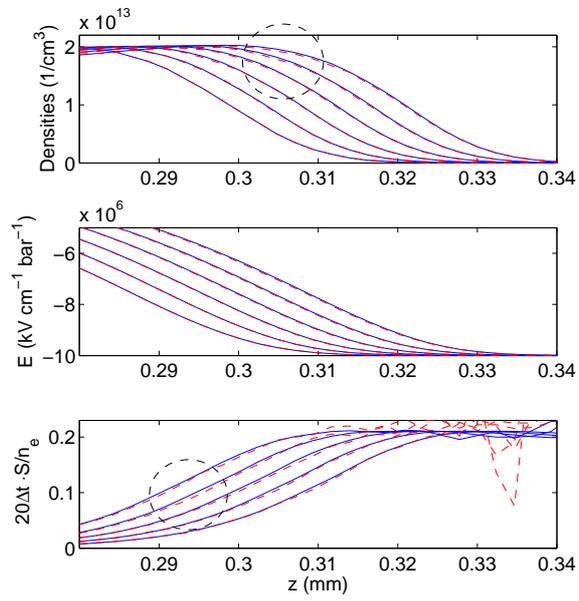}
\end{center}
\caption{\label{fig:drawback_ext} The ionization front propagates into -100 kV/cm in both particle model (dashed) and extend fluid model (solid). The two simulations start from the same front which is pre-produced by a particle simulation, and follow the front propagation for 100~$\Delta t$ where $\Delta t=0.3$ ps. We present the electron density (first panel), the electric field (second panel), and the ionization rate (third panel) in time steps of 20~$\Delta t$.}
\end{figure}

The simulation is done as follows. We first let the particle model follow the evolution of an ionization front until it is fairly smooth. This creates an initial condition that now is run further both with the extended fluid model and with the particle model. The two models are followed for a time of 100~$\Delta t$ where $\Delta t=0.3$ ps in a field of $E^+=-100$~kV/cm. The result is shown in Fig.~\ref{fig:drawback_ext}. In the first panel we plot the electron density profile in steps of 20~$\Delta t$ in the particle model (dashed) and in the extended fluid model (solid). Of course, the curves are identical initially, and a difference builds up in time in the high density region of the front (marked with a dashed circle). In the second panel, we show the electric field which does not vary much between the two models during 100~$\Delta t$. In the third panel, we show the relative growth of the ion density $(\partial_tn_p) /n_e$ integrated over a time of 20~$\Delta t$, which is identical to the local ionization rate ${\cal S}/n_e$ within this time according to Eq.~(\ref{ch6_eq:fluid2}). The figure shows that the ionization rate is slightly higher in the extended fluid model in the region of large $n_e$ (marked with dashed circle), and lower in the region of small $n_e$. Note that these differences appear while the electric fields are the same.
 
The analysis shows that the extended fluid model does not completely reproduce the particle model. This is understandable. The actual ionization rates ${\cal S}/n_e$ depend on the energy distribution of the local electrons. The gradient expansion in the ionization rate relates this energy distribution to the local field and to the electron density gradient. The single adjustable function $k_1(E)$ in this gradient expansion is chosen in such a way that an electron swarm in this field is well fitted. But an electron swarm has a characteristic spatial profile in a given field. 
Regions inside the front might combine a given field with a different density gradient for which the model has not been adjusted. 
A solution would be to allow for more adjustable functions inside the fluid model by expanding in higher order gradients.
In the end, only a fluid model with infinitely many adjustable functions would appropriately describe the averaged behavior of the particle model.
We therefore conclude that the density gradient expansion is a substantial improvement of the fluid model, but that it does not contain the full physics of the particle model.

An alternative improvement of the fluid model next to a gradient expansion was suggested to us recently (see acknowledgement), namely the introduction of the energy equation. 
With the energy equation calculating the local mean electron energies, the fluid model can describe how the electron dynamics depends on the mean local electron energies rather than on the electric field~\cite{Kum1980,Kun1988:1,Guo1993,Boe1995,Kan1998,Hag2005}. 
In appendix~\ref{sec:Energy_equation}, the fluid model with energy equation is used to follow the electron avalanche in a constant electric field. Comparison of Fig.~\ref{fig:swarm_com} and Fig.~\ref{fig:lc_com_en_fp} shows that the extended fluid model follows the full particle avalanche a bit better, but that the fluid model with energy equation fits the mean electron energies in the particle avalanche quite well. 


 We now proceed to the hybrid model where the extended fluid model will be used only in the less critical inner region of the streamer.

\section{The hybrid model}\label{sec:h_model}

 Our goal is to build a fully 3D hybrid model for streamer channels, which couples the particle model with the extended fluid model in space as shown in Fig.~\ref{fig:hybrid_concept}. With the experiments carried out for planar fronts, we would like to test our hybrid concept, and find out how to apply the different models in suitable regions adaptively and how to realize a reasonable coupling of those models. Since most of the electrons will be approximated as densities in the hybrid simulation, the problem of the enormous number of electrons in a 3D particle streamer simulation can be reduced. Following the particles can still be a heavy burden for a hybrid model. Therefore the hybrid simulation will let the fluid model approximate most of the electrons as densities and leave as few electrons as possible for the particle model, under the constraint of producing similar results as the pure particle simulation.

In this section, we present the algorithm for the 1D hybrid calculation and discuss the numerical details. Two important issues of the 1D coupling are discussed in detail: where particle and fluid model are to be applied, and how they should interact with each other.

\subsection{The hybrid algorithm}\label{sec:h_algorithm}

In our hybrid model, the particle model is used in the leading part of the ionization front and the fluid model in the rest of the domain. Between them, we have a model interface. The position of the model interface can be chosen either according to the electron densities or according to the electric field. This means that the model interface is either located at some electron density level $n_e=x~n_{e,max}$ ahead of the electron density peak $n_{e,max}$, or at some electric field level $E=y~ E^+$ where $E^+$ is the electric field ahead the front. In both cases, the real numbers $x,~y$ have to be chosen appropriately within the interval [0,~1].
We discuss this choice and the corresponding position of the model interface in detail in Sect.~\ref{sec:model_interface}.

\begin{figure}
\begin{center}
\includegraphics[width=0.9\textwidth]{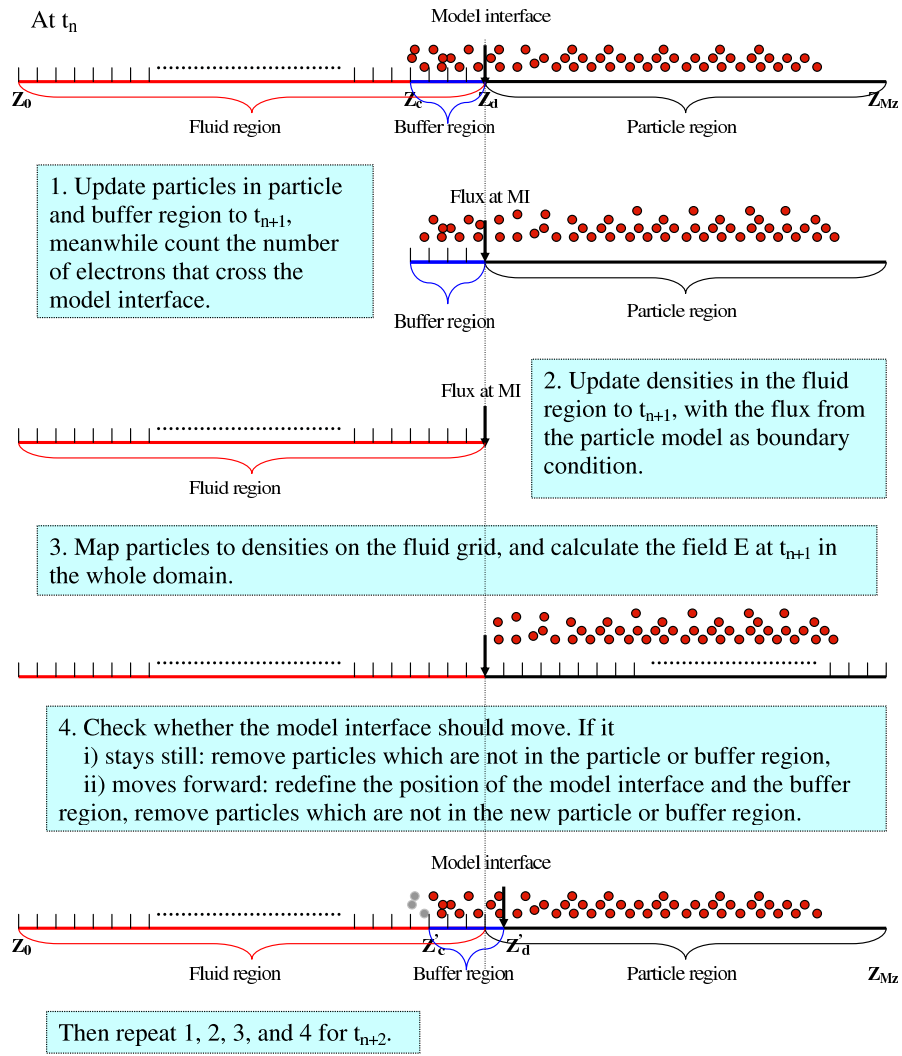}
\end{center}
\caption{\label{fig:hyb_flow} The flow chart of one hybrid computational step from $t_n$ to $t_{n+1}$. }
\end{figure}

Suppose that the position of the model interface is located somewhere as shown in Fig.~\ref{fig:hybrid_concept}. As the streamer propagates forward, because the electric field moves with the front, the model interface moves with the ionization front because we keep it at $n_e=x~n_{e,max}$ or $E=y~ E^+$. In this way, the particle model at any moment of the simulation follows only a limited number of electrons in the low density region of the ionization front. Because the computational costs of the fluid model are small compared to the particle model, it is clear from Fig.~\ref{fig:hybrid_concept} that we gain much efficiency in the hybrid model as compared to a pure particle simulation.

The actual gain due to the hybrid method will be even more pronounced in the case of fully three dimensional simulations. With this moving model interface algorithm, the hybrid model may simulate the full streamer dynamics in 3D without using super-particles while remaining computationally efficient.

Once the position of the model interface is set, the interaction of the two models around the model interface needs to be established. To simulate the electron flux crossing the interface between the two models, we introduced a so called ``buffer region'' which is created by extending the particle region one or a few cells into the fluid region. It has been used in~\cite{Gar1999,Ale2002,Ale2005} for rarefied gases by coupling a Direct Simulation Monte Carlo (DSMC) scheme to the Navier-Stokes equation, and for other applications~\cite{Del2003,Akt2002} as well. The buffer region helps to build a pure particle description around the model interface, such that the local particle flux in this region can be obtained as in a pure particle simulation. The particle flux across the model interface influences the total number of particles in the particle model, and it will lead to a corresponding increase or decrease of the density in the fluid model.

In this way, around the area where the two models are coupled, the electron flux in a pure particle simulation is maintained in the hybrid computation, while global mass conservation holds. However, to obtain an accurate flux at this interface, the electrons in the buffer region near the interface should maintain a correct density profile and velocity distribution. In many cases~\cite{Gar1999,Ale2002,Ale2005}  the buffer region or a part of the buffer region at each time step have to be refilled or reconstructed with a large number of electrons with artificial distributions in energy and space. Such a filling or reconstruction on the one hand ensures a stable flux at the model interface, but on the other hand, it can create non-physical artifacts, which might cause wrong results. In our approach, the generation of electrons from a kinetic prediction can be avoided. The details are presented in Sect.~\ref{sec:buffer_region}.

Fig.~\ref{fig:hyb_flow} shows the flow chart of the hybrid computations. At the beginning of each time step, the particle density is known in the fluid region and the individual particle information is known in the particle region and the buffer region. To update them to the next time step, we first move the particles in the particle and buffer region one step further. The number of electrons crossing the model interface is recorded during the updating. This counted flux is used as a boundary condition to update the densities in the fluid region. To calculate the electric field, the particles in the particle region are mapped to the fluid grid as densities. In the end, we check whether the model interface has to be moved. A more detailed explanation is given in the following section.

\subsection{Numerical implementation}\label{sec:h_implementation}

 In the particle model, the positions and the velocities of the electrons are updated with the leap-frog method. The electric field is solved on a uniform grid $G$ (the fluid equations are discretized on the same grid) with cells \be C_i=\left[ z_i-\frac{1}{2} \Delta z ,  z_i+\frac{1}{2} \Delta z \right], ~i=1,2, \cdots M_z, \ee where $M_z$ are the number of grid points with cell centers at $z_i=(i-1/2) \Delta z$ in the $z$ direction, and $\Delta z$ is the cell size.

The simulation begins with a few electron and ion pairs with a Gaussian density distribution. These initial particles are followed only by the particle model in the beginning of the simulation. As new electrons are generated, the number of particles eventually reaches a given threshold, after which the simulation switches to the hybrid approach. 
The threshold in our simulation is normally set to be several million electrons to ensure a satisfactory statistics. With proper transversal area $A$ of the system, the simulation with such number of electrons is already in the streamer stage, which means that a steady moving ionization front has developed and the charge layer at the streamer head totally screens the field inside the channel. In Table~\ref{tab:system_size_transition}, we summarize the transversal areas ($l_r\times l_r$) and the length of the system in front propagation direction $l_z$ for varying fields, and we also list threshold numbers $N_{T}$ and the time ($T_T$) for the particle simulation generating $N_T$ electrons when starting from 100 pairs of electrons and ions.

\begin{table}
\centering
\begin{minipage}{\textwidth}
\centering
\begin{tabular}{cccccccc}
\hline
\hline
 E (kV/cm) & 50 & 75 & 100 & 125 & 150 & 175 & 200 \\
\hline
$l_r$ ($\mu$m) & 55.2 & 41.4 & 27.6 & 23 & 18.4 & 13.8 & 9.2 \\
$l_z$ (mm) & 6.9 & 3.45 & 2.76 & 2.3 & 1.84 & 1.38 & 1.15 \\
$N_T$ ($10^6$) & 3 & 3.2 & 3.5 & 3.8 & 4 & 5 & 7 \\
$T_T$ (ns) & $\sim$6.6 & $\sim$1.2 & $\sim$0.3 & $\sim$0.18 & $\sim$0.15 & $\sim$0.12 & $\sim$0.12 \\
\hline
\end{tabular}
\end{minipage}
\caption{List of the transversal length $l_r$ (the transversal area is $l_r\times l_r$), the longitudinal length of the system $l_z$, the threshold number $N_T$, and the time $T_T$ for the particle simulation TO generatE $N_T$ electrons for a number of electric fields. }
\label{tab:system_size_transition}
\end{table}

For the transfer from particle to hybrid simulation, we first determine the position of the model interface and the length of the buffer region, which is chosen dependent on the field ahead of the ionization front as we will discuss later. Suppose the model interface is chosen at $z_d=d*\Delta z$ and the buffer region is on the interval $[z_{c},~ z_{d}]$ as shown in Fig.~\ref{fig:hyb_flow}, where $c,d \in \{1,2,\cdots,M_z\}$ and $c \le d$. The particles in the future fluid region in the interval $[z_0,~ z_d]$ are then averaged in this particular time step to the densities on the grid $G$, while the particles outside the particle region and buffer region on the interval $[z_{0},~z_{c}]$ are removed from the particle list. Note that in the buffer region $[z_c,~z_d]$, the electron and ion densities are calculated on the underlying continuum grid while the spatial and kinetic information of the individual particles is also maintained.

Now a general time step in the further evolution is described.
At the beginning of a time step lasting from $t_n$ to $t_{n+1}$, we have the electric field ${\bf E}_n$, the electron and ion densities $n_e$, $n_p$ in the fluid region and the positions of the particles in the particle and buffer region at time $t_n$, and the velocities of these particles at time $t_{n+1/2}$ (since with the leap-frog algorithm of the particle model, in the beginning of the time step, the positions {\bf x} are known at $t_n$ and the velocities {\bf v} at $t_{n+1/2}$). In step 1 (as shown in Fig.~\ref{fig:hyb_flow}), the particles in the particle region and the buffer region are first moved to $t_{n+1}$ taking the stochastic collisions into account through the Monte Carlo algorithm. While updating the particle positions, the number of electrons crossing the model interface is recorded.

Once all particle positions are updated to the next time step, in step 2 we also update the densities in the fluid model. The continuum equations for the particle densities are discretized with a finite volume method, based on mass balances for all cells in the fluid region on the grid $G$. Eq.~\eqref{ch6_eq:flux} is used to compute the density fluxes on the face of each cell in the fluid region, except the one at the model interface $z_d$ where the flux from the particle model is used. The ionization rate {\bf is} calculated with Eq.~\eqref{ch6_eq:S} in the extended fluid model. Given the fluxes on the cell faces and the ionization rate, particle densities at each cell in the fluid region can be updated to the next step using Eq.~\eqref{ch6_eq:fluid1} and Eq.~\eqref{ch6_eq:fluid2} with the nonlocal parameter functions $\mu_n$, ${\bf D}_n$ and $\alpha_n$.

 The densities are calculated using the third order upwind-biased advection scheme. The two-stage Runge-Kutta method could be used with the given particle flux (which is an average flux over $[t_n,~t_{n+1}]$). This will allow computations with larger time step $\Delta t$, but it also requires two updates of the fluid densities and electric fields per time step. Since updating the fluid densities needs the particle flux across the model interface as an input for the boundary condition, the two-stage Runge-Kutta method is replaced by the simpler forward Euler method, as long as the chosen time step is small enough. The time step and grid size in the hybrid simulation are chosen as the same as in the particle simulation and the fluid simulation, see Section~\ref{sec:Swarm_reinvestigation}, i.e., as $\Delta z =2.3 \mu$m and $\Delta t=0.3$ps, therefore the simulation results are compared with each other with the same numerical discretization errors.

Now both the particle positions and the densities are known in the respective regions at time $t_{n+1}$. In step 3, the densities in the particle region $[z_d,~z_{M_z}]$ are obtained by mapping particles to the grid $G$. How this is done, is discussed further below. The particle densities are then known everywhere on the grid $G$ and the electric field $E_z$ at time $t_{n+1}$ can be calculated.

In step 4, the position of the model interface is determined either from the particle density or from the electric field criterion. While the density profile and the electric field are updated from $t_{n}$ to $t_{n+1}$, the model interface may have to move one cell forward or to stay still. With $E^+=-50$ to $-200$ kV/cm, the ionization front needs about 30 to 3 time steps to cross one cell. That is, the model interface will on average stay at one cell face for 3 to 30 time steps before it moves to the next cell face. If it stays still, the particles which fly out of the particle and buffer regions are removed from the particle list.
If the model interface moves one cell forward, the fluid region is extended one cell into the particle region. Meanwhile, the buffer region also moves one cell forward and the region from $z_c$ to $z_c'$ becomes part of the fluid region, and particles there will be removed from the particle list.

Finally we use $E_z$ at $t_{n+1}$ to update the electron velocities inside the particle and buffer region to $t_{n+3/2}$. This finishes one hybrid time step.

We would like to remark here that one should be careful about the technique of mapping the particles to the densities in the particle region. One can use zero-order weighting by simply counting the number of particles within one cell, or first-order weighting (also called Particle In Cell or PIC)~\cite{Bir1991}, which linearly interpolates charges to the neighboring cells, or some higher-order weighting techniques like quadratic or cubic splines. The particle model has been tested with various sizes of time steps and cell sizes. The numerical discretization errors converge to zero as time step and cell size decrease when using either the zero-order weighting or the first-order weighting, but the convergence was faster with first-order weighting. The first-order mapping is the most used technique in particle simulations for plasmas because: ``As a cloud moves through the grid, the first-order weighting contributes to density much more smoothly than zero-order weighting; hence, the resultant plasma density and field will have much less noise and be more acceptable for most plasma simulation problems.''~\cite{Bir1991}. However, in the hybrid computation, the zero-order weighting guarantees total charge conservation in the system, while the first-order weighting may cause charge loss or gain near the model interface when it is applied in a non-uniform density region. Therefore, zero-order weighting is implemented in our hybrid model.

\subsection{The position of the model interface}\label{sec:model_interface}

\begin{figure}
\begin{center}
\includegraphics[width=.95\textwidth]{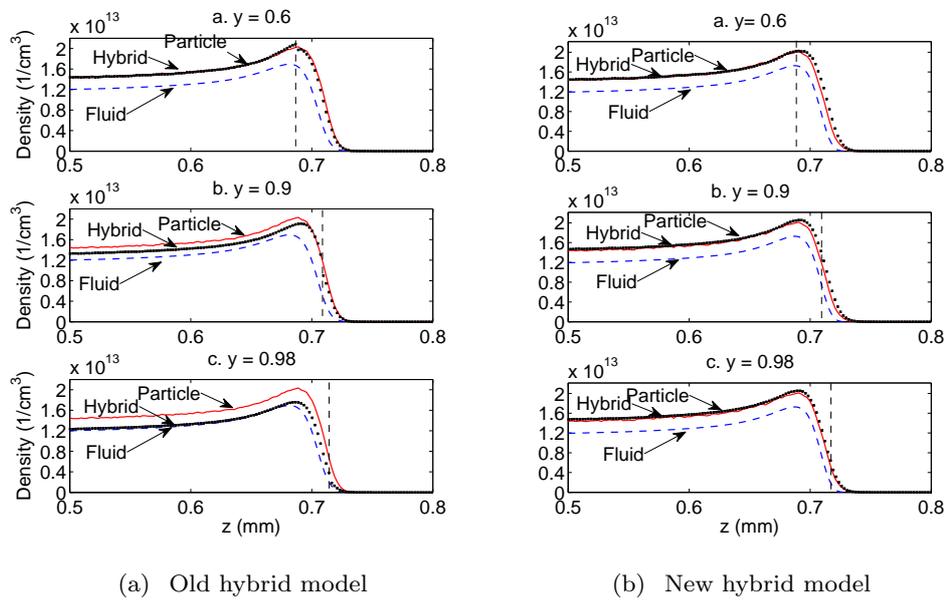}
\end{center}
\caption{\label{fig:fluid_com} The hybrid model (dotted) coupled with the classical fluid model (left column) or the extended fluid model (right column) is compared with the classical fluid model (dashed) and the particle model (solid) for a front propagating into a field of $E^+=-100$ {\rm kV/cm}. The model interface is located at three different levels of the field $E=y E^+$ with $y=0.6$, $0.9$, and $0.98$ shown in the upper, middle and lower panel, respectively.}
 \end{figure}


We have discussed the suitable position of the model interface in~\cite{Li2008:1} where the particle model and the classical fluid model were coupled in space. It was shown that in the front part of the ionization front, the mean electron energies in the classical fluid model were lower than in the particle model, while behind the front the electron energies in both models are in good agreement. Therefore for a good agreement between hybrid and particle simulation, in the hybrid model the fluid model can be used behind the front, and a sufficiently large part of the density decay region should be covered by the particle model.

Now within the hybrid model, the classical fluid model is replaced by the extended fluid model. Since an extra non-local term was introduced, such that the fluid simulation results are now closer to the particle results, we expect that a smaller region needs to be covered by the particle model without changing the result of the hybrid model.

Fig.~\ref{fig:fluid_com} shows the simulation results of an ionization front propagating into a field of $E^+=-100$~kV/cm; the left column shows results of the old hybrid model and the right column those of the new hybrid model. The model interface is located at three different positions: $E=y~ E^+$ where $y=0.6$, $0.9$, and $0.98$, corresponding to the upper, middle and lower panel, respectively. Both in the old and the new hybrid simulation, a long buffer region of 32 $\Delta z$ has been used to ensure the stable interaction of the models at the model interface. On the left, the old hybrid model for $y=0.6$ generates the same electron density behind the front as the particle model, for $y=0.98$ the same density as in the classical fluid model, and for $y=0.9$ some intermediate density; we recall that the density difference between particle and classical fluid model is 20 \% for this field. On the right where the extended fluid model is used, the new hybrid model produces similar results as the particle model in all three cases $y=0.6$, $y=0.9$ and $y=0.98$.

The comparison shows that when the extended fluid model is used instead of the classical fluid model, the performance of the hybrid model is largely improved. Using the extended fluid model in the hybrid computation, the particle model can focus on a smaller portion of the front where the electron density is much lower while the electron density behind the front is still calculated with high accuracy. This greatly reduces the number of electrons that need to be followed in the particle region. It will give a substantial improvement of computational efficiency in a 3D simulation where millions of electrons will be pushed into the fluid region when the fluid model can be applied further ahead within the front.

\begin{figure}
\begin{center}
\includegraphics[width=0.6\textwidth]{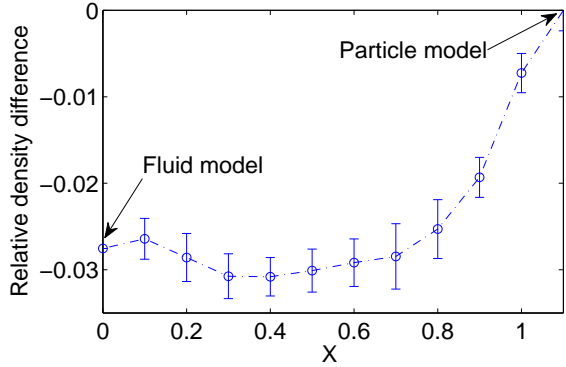}
\end{center}
\caption{\label{fig:err_xmax_100} The relative density differences Eq.~(\ref{EQ-dens}) between the hybrid simulation and the particle simulation as a function of the position of the model interface $n_e=x~n_{e,max}$ for $E^+= -100$~kV/cm, The value $x=0$ accounts for the extended fluid model. The full particle simulation is denoted as $x=1.1$.}
\end{figure}

To determine the proper position of the model interface, different positions have been tested in planar front simulations. Fig.~\ref{fig:err_xmax_100} shows the relative density differences defined in Eq.~(\ref{EQ-dens}) between the hybrid simulation and the particle simulation as a function of the position of the model interface; here the field ahead of the ionization front is $E^+=-100$~kV/cm. The model interface is placed at an electron density level $n_e=x~n_{e,max}$ where $x$ varies. For $x=1$, the model interface is at the density peak $n_{e,max}$ of the ionization front. From $x=1.0$ to $0.1$, the model interface moves forward within the front, and $x=0$ corresponds to a full fluid simulation. The full particle simulation is denoted as $x=1.1$.

\subsection{The buffer region}\label{sec:buffer_region}

The particle model extends backward, beyond the model interface, into the buffer region where particle and fluid model coexist; it supplies particle fluxes for the fluid model at the model interface, as illustrated in Fig~\ref{fig:hyb_flow}. However, correct particle fluxes require correct particle statistics within the buffer region whose length should be as small as possible to reduce computation costs, but larger than the electron energy relaxation length~\cite{Li2007}.

\subsubsection{Adding particles in the buffer region}

As we have mentioned, to ensure a stable electron flux at the model interface, in general new particles need to be introduced into the buffer region, that have to be drawn from appropriate distributions in configuration space. This would pose a particular problem for the streamer simulation, since Maxwellian or even Druyvesteyn~\cite{Dru1940} distributions are inaccurate. However, even for negative streamers, the electrons on average move somewhat slower than the whole ionization front, which means that the electrons on average are moving from the particle region into the fluid region.

This is because in a pulled negative ionization front, the front velocity is determined in the leading edge of the front where the electric field and therefore the electron drift velocity is highest. But the front velocity is determined by this maximal electron velocity plus an additional positive contribution of electron diffusion and impact ionization~\cite{Ebe1997}, therefore the front everywhere is faster than the local mean electron velocity.

Suppose that when the computation is changing from the pure particle regime to the hybrid regime, we create a buffer region which is long enough to relax most of the fast electrons. The particles in the buffer region are the heritage of the pure particle simulation and will keep being followed by the particle model. Across one end of the buffer region, the model interface at $z_d$, particles move freely. But across the other end of the buffer region at $z_c$, there are electrons flying out of the buffer cells but no electrons enter, i.e., when the model interface stays still, the buffer cells are losing particles. If we add electrons in the buffer region near $z_c$ to create a flux from the fluid cell $z_{c-1}$ into the buffer region $z_c$,  although the particle loss is compensated, these inflowing particles have hardly any influence on the flux across the model interface when $z_d-z_c$ is sufficient long. Once the model interface moves forward, those added particles near $z_c$ will fall into the fluid region and be removed again.

This hypothesis has been tested in hybrid simulations with two kind of fluxes used at $z_{c-1/2}$:
\begin{itemize}
\item No influx: electrons can only fly out over $z_{c-1/2}$, but not fly back.
\item Reflected influx: electrons that fly out over $z_{c-1/2}$ fly back immediately with inverted velocity.
\end{itemize}
The hybrid simulation has been carried out for a planar front at $-100$~kV/cm with the model interface at $E=0.6$, 0.8, 0.9  E$^+$. With a long buffer region of 10 $\Delta z$, there is no influence of the inflowing particles. We even tested the artificial case of "double reflection influx", where twice as many electrons fly back with doubled energies. Even then, there was no notable influence.

So we conclude that if electrons move on average more slowly than the front, the electron loss at the end of
a sufficiently long buffer region does not affect the calculation of particle fluxes at the model interface.
Therefore the particles lost at the end of the buffer region can be ignored and new electrons do not need to
be created artificially. This actually leads to a simpler technique of model coupling in which the correct energy distribution of particles is not a concern any more. This technique is not only suitable for streamer simulations, but for all two phase problems where the front speed is higher than the speed of the individual particles, as is generally the case in ``pulled front'' problems~\cite{Ebe2000:2,Saa2003}, cf.~section~\ref{sec13}. For example, bacterial growth and transport, or wound healing as a front propagation problem can be treated in this way.

\begin{figure}
\begin{center}
\includegraphics[width=0.48\textwidth]{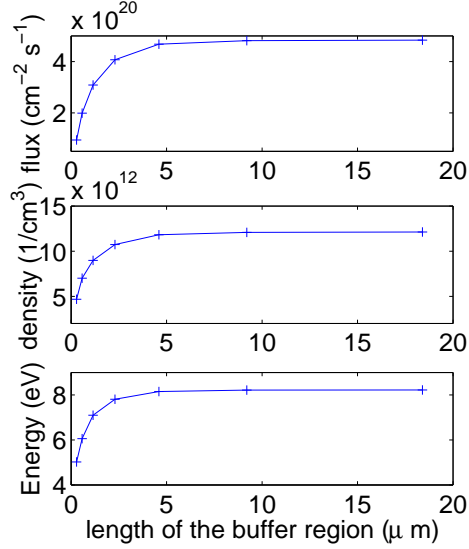}
\end{center}
\caption{\label{fig:buff_conv_90} Hybrid calculation of an ionization front propagating into the field of $-100$ kV/cm with the model interface placed at 0.9 $E^+$. The plots show the average electron flux (first panel) {\bf across} the model interface, and densities (second panel) and mean energies (third panel) of the electrons {\bf at} the model interface as a function of the buffer length region. It shows that all the quantities converge when buffer length increases.}
\end{figure}

\subsubsection{The length of the buffer region}\label{sec:Ch6_length_of_buffer}

A reasonable particle flux can be obtained without artificially adding new electrons in a sufficiently long buffer region. But how long is enough long?
Although electrons on average propagate slower than the model interface, the high energy electrons at the tail of the energy distribution may move faster than the ionization front in a short time interval. The buffer region should be long enough that most of the fast electrons can relax to mean energies within this short time.
The relaxation length of the fast electrons at the ionization front has been discussed in~\cite{Li2007}.
It shows that three electron swarms with identical energies of 0.5, 5, or 50 eV, respectively, equilibrate  within 2 ps through losing or gaining energy in collisions.
If the buffer region is very short, high energy electrons might enter from the fluid region and contribute to the flux on the model interface, but since they are not be included in the buffer region, such short buffer region will result in a lower flux on the model interface. 

A stable flux can be obtained only when the buffer region is long enough to relax most high energy electrons to the local energies within the buffer region.
An upper bound for the mean forward velocity ($v_z$) of local electrons is the front velocity $v_f \approx 7\times 10^5$ m/s at -$100$ kV/cm.
When an electron with 50 eV and with very small radial velocity components, i.e., with $v_z \gg v_r$, relaxes to the local electron energy, its velocity in the forward direction $v_z'$ decreases from $\sim 4\times 10^6$ m/s to $v_f$ within 2 ps, and it propagates $2\times10^{-12} \cdot (v_z'+v_f)/2 =4.7$ $\mu$m, which is approximately 2 cells. 

Experiments with different lengths of the buffer region are shown in Fig.~\ref{fig:buff_conv_90}. Here we plot average flux, density and mean energy of the electrons at the model interface for buffer regions of different lengths. The field ahead the front is $E^+= -100$ kV/cm and the position of the model interface is at $E=0.9 E^+$. 
It is clear that as the length of the buffer region increases, not only the flux densities, but also the electron density and the mean electron energy converge to their limit values. Our computational cells are 2.3$\mu$m long, and a buffer region with the length of 2 cells can already give a reasonable flux at the model interface. 

\section{Simulation results in different fields}\label{sec:Ch6_other_field}

Having presented our hybrid simulations of the front propagating into a field of $E^+=-100$ kV/cm in detail, we now summarize results for fields ranging from $-50$ to $-200$ kV/cm.

\begin{figure}
\begin{center}
\includegraphics[width=0.6\textwidth]{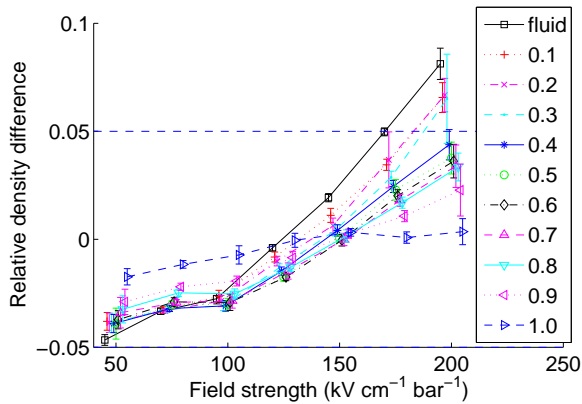}
\end{center}
\caption{\label{fig:err_xmax} The relative density differences between the hybrid simulation and the particle simulation as a function of the position of the model interface in a range of fields. The model interface is at the position where the electron density level is $n_e=x~n_{e,max}$. }\end{figure}

In Fig.~\ref{fig:err_xmax}, we present the relative density discrepancies $\left(n_{e,part}^- -
n_{e,hybr}^-\right)/{n_{e,part}^-}$ of the saturated electron density $n_e^-$ behind the ionization front for different fields; here $hybr$ denotes hybrid simulations and $part$ particle simulations. For each field, the model interface has been placed at electron density levels $n_e=x~n_{e,max}$, with $x=1.0$, $0.9$, $\dots$, $0.1$. 
A long buffer region with 32 cells was used in the hybrid calculation to ensure a stable flux at the model interface. Two horizontal dashed lines have been added at $\pm 5\%$ to assist the choice of the proper position of the model interface. For example for a field of $E^+= -125$~kV/cm, the relative density difference is limited to $\pm 2\%$ even if only a very small part of the front $x=0.1$ is covered by the particle model. But due to the flux convergence problem discussed in Sect.~\ref{sec:buffer_region}, the model interface will be put slightly more backwards to have a stable flux in our actual hybrid computational model, as the computational costs are actually determined by the position of the back end of the buffer region.

\begin{table}
\begin{center}
\begin{tabular}{|c|c|c|c|c|c|c|c|}
    \hline
    \hline
$~~E^+~~$  & 50 & 75 & 100 & 125 & 150 & 175 & 200 \\
$~~~x~~~$  & 0.5-0.9 & 0.55-1.0 & 0.6-1.0 & 0.65-1.0 & 0.7-1.0 & 0.8-1.0 & 0.9-1.0  \\
tested $x$  & 0.5 & 0.55 & 0.6 & 0.65 & 0.7 & 0.8 & 0.9  \\
buffer cells  & 1 & 1 & 2 & 2 & 2 & 3 & 3  \\
    \hline
\end{tabular}
\caption{\label{tab:mi} The proper position of the model interfaces for different fields, where $n_e=x~n_{e,max}$.}
\end{center}
 \end{table}

We summarize the proper positions of the model interfaces for different fields in Table.~\ref{tab:mi}.
With the positions chosen from this table, we have tested different lengths of buffer regions.
For the buffer lengths given, the electron flux, density and mean energy lie not more than $2\%$ off the limit of a very long buffer region.

\section{Conclusion}\label{sec:ch6_conclusion}

The particle model contains all essential microscopic physical mechanisms of a streamer ionization front, but the computational effort increases with the growing number of particles and rapidly exceeds computer memory. Gathering many real particles into one super-particle is not an option, as it causes wrong statistics and causes numerical artifacts. The fluid model is much more efficient than the particle model since the particles are approximated as continuous densities, but particle density fluctuations and large deviations of individual electron energies from the mean are not modeled. To combine the computational efficiency of the fluid model and the full physics of the particle model, a spatially hybrid model has been implemented for planar fronts. The hybrid model uses a particle model in the leading part of the ionization front where the total electron number is low and the field is high, and an extended fluid model for the rest. Once the scheme is extended to 3D, it can represent particle number fluctuations in the leading edge that might play a role in accelerating the branching process, or it can follow the run-away of individual electrons from the front as a possible cause of X-ray bursts and gamma-ray flashes.

For constructing the hybrid model, the fluid model has been extended with a density gradient term in the impact ionization term. This extended fluid model has been introduced as an alternative to the classical fluid model, since it maintains an electron flux that is consistent with the particle model. Both avalanche experiments and planar front experiments confirm that the extended fluid model should be used in the hybrid calculation rather than the classical fluid model. (The fluid model with energy equation is discussed in appendix~\ref{sec:Energy_equation} as a possible alternative.)

The hybrid algorithms and our numerical implementation are discussed in detail, in particular, the position of the interface between particle and fluid model, and the construction of the buffer region at the interface. After testing the hybrid model at $-100$~kV/cm, we also present the hybrid simulation results for other fields, resulting in recommendations for the position of the model interface and the length of the buffer region in future 3D simulations.

Although the spatially hybrid model is discussed in the context of streamer simulations, we would like to emphasize two particular points that are generic for pulled front problems. 1. When particle avalanches are used to calculate the transport and reaction coefficients for the fluid model, particle or fluid fronts agree particularly well, as they have the same front shape as a particle avalanche; this property holds only for pulled fronts and not for pushed or bistable fronts. 2. As the electrons on average move backward in a frame moving with the front velocity, the problem of creating particles with proper statistics in the buffer region does not need to be solved. This problem here would be particular severe, as the distribution of electron energies and velocities are far from thermal and not even in equilibrium with the local electric field. But in many other cases, the distribution of inflowing particles is unavailable or difficult to obtain as well. The simplified coupling approach here offers an alternative for those problems without having to consider the influx on the back end of the buffer region. 
\\

{\bf Acknowledgment:} The authors would like to M. Kushner, L. Pitchford, J.J.A.M. van der Muller and J. van Dijk for participating in the Ph.D. thesis defense committe of C. Li, and for suggesting the use of the energy equation. The authors acknowledge the support of the Dutch National
Programe BSIK, in the ICT project BRICKS, theme MSV1. C.L. also acknowledges recent support
in the STW-project 10118 of the Netherlands' Organization for Scientific Research NWO.

\newpage
\begin{appendix}
\section{The electron energy equation}\label{sec:Energy_equation}

A set of moment equations can be derived from the Boltzmann equation, in which the most significant quantities are the density, momentum, and energy of the electrons. The Boltzmann equation is of the form
\be \label{equ:Boltzmann}   \frac{\partial f}{\partial t} + {\bf v} \cdot \nabla_r f + \frac{\bf F}{m}\cdot \nabla_{\bf v} f = \left( \frac{\partial f}{\partial t} \right)_c ,  \ee
where $f({\bf r},{\bf v},t)$ is the distribution function of the considered charged particle depending on time t, position {\bf r} and velocity {\bf v}, $m$ is the particle mass, {\bf F} is the electrical force exerted on the particle, and the term in the right hand side of Eq.~(\ref{equ:Boltzmann}) is the time rate of change of $f$ due to collisions. 

The evolution of electron density, electron average momentum and electron mean energy are defined by the zero-, first-, and second-order moments of the distribution function~\cite{Mit1973,Shk1966}. 
For example, the density equations in the classical fluid model, Eq.~(\ref{ch6_eq:fluid1}) and Eq.~(\ref{ch6_eq:fluid2}), can be deduced from the Boltzmann equation by integrating Eq.~(\ref{equ:Boltzmann}) over $d${\bf v}. 
The momentum equation can be obtained by multiplying Eq.~(\ref{equ:Boltzmann}) by $m${\bf v} and integrating over $d${\bf v};
The energy equation can be obtained by multiplying Eq.~(\ref{equ:Boltzmann}) by $\frac{m}{2}$~{\bf v}$\cdot${\bf v} and integrating over $d${\bf v}.  

Here we will mainly discuss the electron energy equation as a potential solution for the non-local effects reported in this paper and in~\cite{Li2007}. 
We have shown in~\cite{Li2007} that in the particle simulation, the mean energies of local electrons at the front are higher than in the classical fluid simulation, both in the electron avalanche and in the planar front, see figures $7$ and $10$ in~\cite{Li2007}. 
In Section~\ref{sec:Swarm_reinvestigation}, we concluded that the local field and the local density approximation are not sufficient to describe the non-local spatial energy distribution appearing in the particle simulation, which causes the flux discrepancy between the particle and fluid description in a hybrid calculation, see Section~\ref{sec:why_ex}.
This problem is addressed by an extended fluid model which approximates the non-equilibrium ionization rate with a density gradient term, see Section~\ref{sec:f_equations}. 
However, since the electron behavior (drift, diffusion, collision frequency) is determined by the electron energies, that are approximated by the local electric field, the equation for the local mean electron energy is here explored as an alternative to the gradient expansion. 
The energy equation can describe the non-equilibrium behavior of electrons in an electric field if $i$) the transport and reaction coefficients can be approximated as functions of the mean electron energies, 
and $ii$) the electron mean energies are calculated with an electron energy equation. 

We first review the transport coefficients and redefine the electron flux ${\bf j}_e $ and ionization source term ${\cal S} $ in the density equations (\ref{ch6_eq:fluid1}) and (\ref{ch6_eq:fluid2})
\ba {\bf j}_e &=& - \mu(\bar{\epsilon})~ {\bf E}~ n_e - {\bf D}({\bar{\epsilon}}) \cdot{\bf \nabla} n_e, \label{eq:energy_fluid_flux} \\
{\cal S} &=& |n_e\;\mu(\bar{\epsilon})\;{\bf E}|\;\alpha(\bar{\epsilon}),
\ea
where the transport coefficients $\mu$ and ${\bf D}$, and the ionization rate $\alpha$ are now functions of the mean electron energy $\bar{\epsilon}$.

The electron energy equation has been used in various plasma simulations, for example, for rf (radio frequence) discharges~\cite{Boe1995,Rai1995,Ham1999,Dav2009}, for lamp breakdown processes~\cite{Bho2004}, and also for streamers~\cite{Bay1985,Guo1993,Kan1998}. 
The equation has the general form:
\be \label{equ:energy_equation_ge} \frac{\partial(n_e \bar{\epsilon})}{\partial t} +\nabla \cdot {\bf j}_{{\epsilon}} = - e {\bf j}_e \cdot {\bf E}+ \left(\frac{\partial(n_e {\epsilon})}{\partial t}\right)_c  ,  \ee
where ${\bf j}_{{\epsilon}}$ is the energy flux (not to be confused with ${\bf j}_e$), and $(\frac{\partial(n_e {\epsilon})}{\partial t})_c$ represents the total energy loss due to collisions. 

The energy flux can be written in drift-diffusion form similarly to the density flux as 
\be {\bf j}_{\epsilon} = - \mu_{\epsilon}(\bar{\epsilon})~{\bf E}~ n_e~ \bar{\epsilon}  - {\bf D}_{\epsilon}({\bar{\epsilon}})~ {\bf \nabla} n_e  \bar{\epsilon}  \label{eq:energy_fluid_flux_1}, \ee
where the $ \mu_{\epsilon}$ and ${\bf D}_{\epsilon}$ are the energy mobility and the energy diffusion coefficients. Many fluid models in the literature~\cite{Bay1985,Boe1995,Rai1995,Ham1999,Bho2004,Dav2009} use the energy transport coefficients given by 
\be \mu_{\epsilon}= \frac{5}{3}~ \mu \hspace{0.5cm} \mbox{and} \hspace{0.5cm}  {\bf D}_{\epsilon}= \frac{5}{3}~ {\bf D}. \label{eq:energe_den_coe}\ee These approximations can be derived by assuming a Maxwellian EEDF, a constant momentum-transfer frequency and constant kinetic pressure~\cite{Kan1998,Hag2005}.   

Concerning the energy collision term, the rate coefficients can be directly obtained from electron swarm experiments. The rate coefficients for excitational and ionization collisions are calculated in the form of Townsend coefficients $\alpha_k(\epsilon)$, where $k$ stands for the $k$th inelastical collision type with the energy loss $\epsilon_{k}$. If we have $m$ electron-neutral inelastical collision processes, the time rate of change of electron energy due to collisions is the following:
\be  \left(\frac{\partial(n_e {\epsilon})}{\partial t}\right)_c =  \sum_{k=1}^{m}~ |{\bf j}_e|~ \alpha_k~ \epsilon_{k}  \ee
A practical form of the energy equation is, 
\be \label{equ:energy_equation_pra} \frac{\partial(n_e \bar{\epsilon})}{\partial t} +\frac{5}{3}\nabla \cdot ({\bf j}_e ~ {\bar{\epsilon}}) = - e~ {\bf j}_e \cdot {\bf E} - \sum_{k=1}^{m} |{\bf j}_e|~ \alpha_k ~\epsilon_{k} .  \ee 

\begin{figure}
\begin{center}
\includegraphics[width=0.7\textwidth]{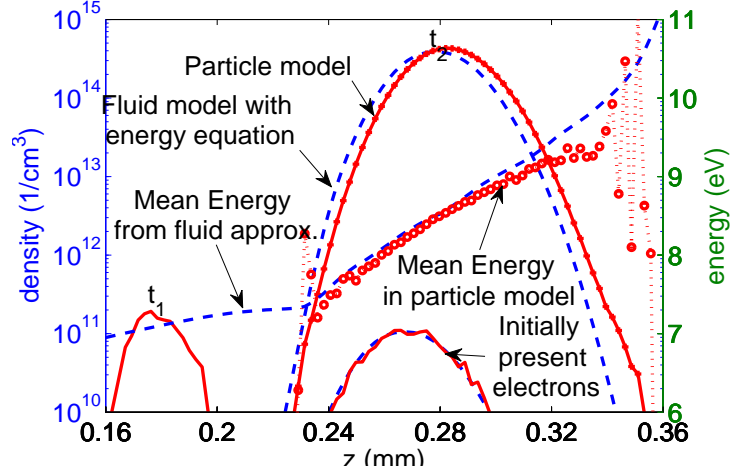}
\end{center}
\caption{\label{fig:lc_com_en_fp} Data of the particle model and of the fluid model with energy equation are shown: the density profile of the electron swarms and the initial present electrons in a field of -100 kV/cm are presented at two different time $t_1$ and $t_2$; we also show the mean energies of electrons of swarms at time $t_2$.} 
\end{figure}

The energy equation coupled with the density equation is then solved for an electron avalanche. The simulation results are presented in Fig.~\ref{fig:lc_com_en_fp}. 
As in Section~\ref{sec:Swarm_reinvestigation} and Fig.~\ref{fig:swarm_com}, a small electron swarm is first generated by the particle simulation at $t=t_1$, then fluid model and particle model follow it to $t=t_2$. Plotted are the same quantities as in Fig.~\ref{fig:swarm_com} plus fluid and particle results for the local mean electron energies at time $t_2$, the particle results for the electron energy were first reported in Fig.~$10$ in~\cite{Li2007}.
 


As shown in Fig.~\ref{fig:lc_com_en_fp}, the fluid model with energy equation generates a similar swarm as the particle model while small differences remain in velocity and maximum electron density. 
The local energy description in particle simulation and fluid simulation agrees surprisingly well. 
Note that this nice agreement is obtained when the energy flux coefficients adopt the commonly used approximation as shown in Eq.~(\ref{eq:energe_den_coe}), where the electron energy distribution function is assumed to be Maxwellian, though previous studies have shown that the electron energy distribution function is not Maxwellian during an avalanche in the high field, cf. Fig.~$3$ in~\cite{Li2007}.  

The energy equation is an alternative for the gradient expansion or vice versa, since both methods can describe the deviation of electrons from the local field approximation. The deviation has been addressed as the `non-local' effect in the density and field gradient approach~\cite{Nai1997}, or as the `non-equilibrium' effect in the multi-moment equation approach~\cite{Kun1988:1,Guo1993,Kan1998}, which shows the two different perspectives of these two approaches. 

Which approach is better in the hybrid model? The accuracy of the coefficients plays an essential role. The electron `non-local' rate coefficients $k_i$ in Eq.~(\ref{ch6_eq:Ale}), including all combinations of the density gradients and field gradients, have been calculated by solving the Boltzmann equation in a two-term approximation~\cite{Ale1996}. 
Some authors working with the energy equation~\cite{Ing1997,Alv1997,Hag2005} also avoid the commonly used energy transport coefficients approximation in Eq.~(\ref{eq:energe_den_coe}), and calculate them more precisely using the two-term Boltzmann equation approximation.
However, it is not clear, how consistent the calculated coefficients from the two-term Boltzmann equation are with the particle description, while consistency is important for the hybrid coupling. 
When the classical fluid model is extended with the first and most important term in Eq.~(\ref{ch6_eq:Ale}), the density gradient, the rate coefficient of this term can be easily calculated from the particle swarm experiment. Since this simple extension already brings a good agreement in electron flux and since it is sufficient for the hybrid coupling, more gradient terms or the energy equation are not introduced in the fluid model in this paper.

\end{appendix}


\end{document}